\documentclass[twocolumn,showpacs,preprintnumbers,prl]{revtex4}
\usepackage{graphicx}
\usepackage{dcolumn}
\usepackage{amssymb}
\usepackage{bm}

\begin{document}

\title{Striped quantum Hall state in a half-filled Landau level}
\author{Xin Wan}
\affiliation{Zhejiang Institute of Modern Physics, Zhejiang University,
Hangzhou 310027, China\\
Collaborative Innovation Center of Advanced Microstructures,
Nanjing 210093, China}
\author{Kun Yang}
\affiliation{National High Magnetic Field Laboratory and Department of Physics,
Florida State University, Tallahassee, Florida 32306, USA\\
Department of Physics, Tsinghua University, Beijing 100084, China}

\begin{abstract}
Nature of the fractional quantum Hall state at Landau level filling
factor 5/2 remains elusive despite intensive experimental and
theoretical work. While the leading theoretical candidates are
Moore-Read Pfaffian (Pf) and its particle-hole conjugate anti-Pfaffian
(APf), neither received unambiguous experimental support. We show that
a state that is intermediate between them, made of alternating stripes
of Pf and APf in the bulk, is a viable candidate. Such a state is
shown to be incompressible and thus a charge insulator in the bulk,
but a heat conductor due to the presence of gapless neutral bulk
modes. We argue that properties of such a state is consistent with
existing numerical and experimental work, and discuss possible
experimental probes of its presence.
\end{abstract}

\date{\today}

\maketitle

The fractional quantum Hall (FQH) state at Landau level (LL) filling
factor $\nu=5/2$~\cite{willett87} has been one of the main focuses in
experimental and theoretical studies of FQH effect for well over a
decade. The intense interest in this state lies in its likely
non-Abelian nature and potential application for topological quantum
computation~\cite{nayak08}. Leading theoretical candidates include
Moore-Read Pfaffian (Pf) state~\cite{moore91}, which is the very
first example of a non-Abelian state of matter, and its closely
related particle-hole conjugate, anti-Pfaffian (APf)
state~\cite{levin07,lee07}. Determining the nature of the 5/2 FQH
state experimentally would not only resolve a long-standing issue in
the field of quantum Hall effect, but also be of tremendous interest
to the physics community in general, in particular if the outcome is
an unambiguous demonstration of a non-Abelian topological state of
matter for the first time.

The Pf and APf states have identical bulk excitation spectra, and
cannot be distinguished by bulk thermoelectric and thermodynamical
probes~\cite{yanghalperin,cooperstern,gervaisyang,barlasyang} (result
of an initial attempt appears to be consistent with
both~\cite{Chickering12}). Qualitative differences are in
their edge properties, which in principle allow for distinction between them as well
as other candidate states through electron and quasiparticle
tunneling. Recent experiments~\cite{radu08,dolev11,lin12} appear to be
more consistent with the APf state, but cannot rule out the Pf state due to possible
edge reconstruction~\cite{overbosch08} or the 331 and other Abelian
states~\cite{lin12,yang13}.  Numerically, Pf and APf states are exactly
degenerate in energy in the absence of LL mixing due to particle-hole
symmetry for a half-filled LL. LL mixing breaks particle-hole symmetry
and is expected to lift this degeneracy, but it remains controversial
which state would be favored energetically at this
point~\cite{wojs10,rezayi11,papic12,pakrouski15,zaletel15,tylantyler15}.

\begin{figure}
\includegraphics[width=8cm]{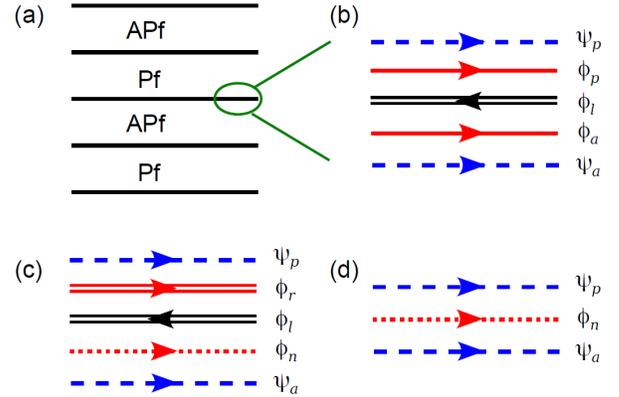}
\caption{\label{fig:domain} (Color online) (a) A striped phase
  containing alternating domains of Pf and APf states.  (b) Original
  edge modes along a Pf-APf domain wall, containing three bosonic
  modes $\phi_p$, $\phi_a$, $\phi_l$, and two fermionic modes $\psi_p$
  and $\psi_a$. (c) The symmetric and antisymmetric linear
  combinations of the two $\nu = 1/2$ bosonic modes $\phi_p$ and
  $\phi_a$ form a charge mode $\phi_r$ ($\nu^{*} = 1$) and a neutral
  mode $\phi_n$. (d) Electron pair tunneling (see Fig. \ref{fig:pair})
  along with strong Coulomb interaction can gap the counterpropagating
  charge modes $\phi_l$ and $\phi_r$, leaving the bosonic neutral mode
  $\phi_n$ and two neutral fermionic modes $\psi_p$ and $\psi_a$ to
  remain gapless. See text for details.}
\end{figure}

The purpose of the present work is to point out that a new state made
of alternating stripes of Pf and APf states in the bulk, is a possible
candidate of the FQH state at 5/2 [see Fig.~\ref{fig:domain}(a) for an
  illustration]. We show that such a striped state, unlike those
realized in high LLs~\cite{fogler96,moessner96,lilly99,du99,rhy}, is
{\em incompressible} and thus a FQH state. On the other hand it
supports {\em gapless} neutral modes in the bulk, and thus has the
novel property of being a bulk heat conductor. We argue that its
properties are qualitatively consistent with existing experiments and
numerical studies, and suggest new experiments to probe its presence.

While the FQH state at 5/2 was discovered experimentally in
1987~\cite{willett87} and the Pf state first written down in
1991~\cite{moore91}, the connection between them started to be taken
seriously only after numerical work by Morf in 1998~\cite{morf98} and
Rezayi and Haldane (RH) in 2000~\cite{rezayi00}. At the time of these
seminal numerical works the APf state was not known. In Morf's work a
shift that specifically favors the Pf state was chosen on the sphere, while
in that of RH the Pf state was particle-hole symmetrized by hand on
the torus when comparing with numerical ground state obtained by
exactly diagonalizing the Coulomb Hamiltonian of a half-filled
first-excited LL. Intriguingly, RH also found that the (particle-hole
symmetrized) Pf state is actually stable only in a very narrow region
in their phase diagram, sandwiched between the (compressible)
composite fermion Fermi liquid-like state and a striped state. With the
additional insight of the existence of the APf state, we now propose that
the striped state is nothing but a mixture made of alternating stripes
of Pf and APf states. It thus appears very naturally in the phase
diagram very close to the (particle-hole symmetrized) Pf state. More
importantly, the incompressible nature (as we will demonstrate below)
of this striped FQH state explains the robustness of the
experimentally observed FQH state at 5/2, despite the numerical
finding of RH that the particle-hole symmetrized Pf state is stable
only in a narrow region of parameter space.

{\em Domain wall between Pf and APf states} --- Pf and
APf states are both incompressible. The striped state [as illustrated
  in Fig.~\ref{fig:domain}(a)] contains domain walls separating Pf and
APf states. Such domain walls are expected to support {\em gapless}
modes, just like edges. One thus might expect there are gapless charge
modes in the bulk due to the presence of these domain walls, and the
state would be compressible (just like striped states in higher
LLs). We now show that in the present case the domain wall only
supports neutral modes but the charge modes are gapped due to presence
of electron tunneling between different modes and strong Coulomb
interaction. As a result we arrive at a novel incompressible state
that supports gapless neutral modes in the bulk.

The Lagrangian density of the Pf-APf interface includes terms
corresponding to bosonic and fermionic modes~\cite{milovanovic96}:
\begin{eqnarray}
L&=&\frac{1}{4\pi}[\partial_t\phi_l \partial_x\phi_l
-2 (\partial_t\phi_p \partial_x\phi_p + \partial_t\phi_a \partial_x\phi_a)]
\nonumber \\
\label{eq:dynamical terms}
&-&[i\psi_p\partial_t\psi_p + i\psi_a\partial_t\psi_a]
 - H[\phi,\psi],
\end{eqnarray}
where $\phi_l$ is the (left moving) bosonic edge field of the $\nu=1$
background in which the embedded holes forming the Pf state (thus
resulting in the APf state), $\phi_p$ and $\phi_a$ are the edge
bosonic fields of the Pf state for the electrons and holes
respectively, and $\psi_p$ and $\psi_a$ are the corresponding edge
Majorana fermion fields, as illustrated in
Fig.~\ref{fig:domain}(b). $H$ is the Hamiltonian density.

The charge density along the interface
\begin{eqnarray}
\rho(x)=\frac{1}{2\pi} \partial_x(\phi_l+\phi_p + \phi_a)
=\frac{1}{2\pi} \partial_x\phi_c,
\end{eqnarray}
where
\begin{eqnarray}
\phi_c=\phi_l+\phi_p + \phi_a
\end{eqnarray}
is the total charge field.
This naturally suggests the following combination of $\phi_p$ and $ \phi_a$:
\begin{eqnarray}
\phi_{r,n}=\phi_p \pm \phi_a,
\end{eqnarray}
where $\phi_r$ is a right-moving charge field and $\phi_n$ is a
right-moving {\em neutral} bosonic field. In terms of these
combinations the dynamical terms involving bosonic fields in $L$
[Eq. (\ref{eq:dynamical terms})] become
\begin{eqnarray}
\frac{1}{4\pi}[(\partial_t\phi_l \partial_x\phi_l -
\partial_t\phi_r \partial_x\phi_r) - \partial_t\phi_n \partial_x\phi_n],
\label{eq:new dynamical terms}
\end{eqnarray}
as illustrated in Fig.~\ref{fig:domain}(c).
Note in particular the two terms combined in the bracket above is the
same as those of left and right movers in an ordinary (or non-chiral)
Luttinger liquid.

All terms allowed by symmetry show up in $H$ with varying
magnitudes. Of particular importance to the physics we discuss here
are the following.

\begin{figure}
\includegraphics[width=7.5cm]{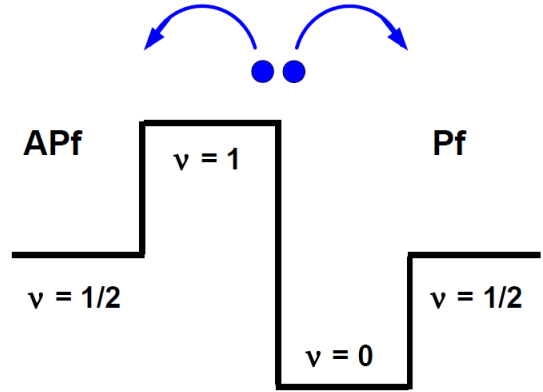}
\caption{\label{fig:pair} (Color online) The relevant
  momentum-conserving tunneling process. A pair of electrons tunnel
  from the $\nu=1$ edge into the Pf edges of electrons and
  holes, respectively. }
\end{figure}

(i) A pair of electrons tunnel from the $\nu=1$ edge into the Pf
edges of electrons and holes (see Fig.~\ref{fig:pair}), respectively,
and its hermitian conjugate (h.c.):
\begin{eqnarray}
T_p\propto \psi_p\psi_a e^{2i(\phi_l+\phi_p+\phi_a)}+ h.c.
= 2\psi_p\psi_a \cos(2\phi_c).
\end{eqnarray}

(ii) Strong Coulomb interaction
\begin{eqnarray}
V_c=v_c(\partial_x\phi_c)^2.
\end{eqnarray}
In the absence of screening (say due to nearby metallic gates) $v_c$
diverges logarithmically in the long-distance limit, due to the
long-range nature of Coulomb interaction. In practice this renders
$v_c$ much larger than kinetic energy and other interaction terms
involving neutral fields. This significantly reduces the scaling
dimension of $\cos(2\phi_c)$, making it
\begin{eqnarray}
\Delta_c\ll 1.
\end{eqnarray}
As a result of this we expect the scaling dimension of $T_p$
\begin{eqnarray}
\Delta_{T_p}=2\Delta_\psi + \Delta_c = 1 + \Delta_c < 2.
\end{eqnarray}
This means $T_p$ is a {\em relevant} perturbation under
renormalization group, and develops an expectation value via the usual
mechanism as in the sine-Gorden model (or in ordinary Luttinger liquid
with back-scattering). This opens up a gap for the charge modes
$\phi_l$ and $\phi_r$, leaving us with a single gapless neutral
bosonic mode $\phi_n$ and two fermionic modes, as illustrated in
Fig.~\ref{fig:domain}(d). Note that they are propagating along the
same direction and the total central charge is $c = 1 + 1/2 + 1/2 =
2$.

{\it Numerical evidence.} Now we revisit the RH work in sphere and
torus geometries. The study showed that there is a first-order
transition from the symmetrized Pf state to a compressible striped
phase. The single Slater determinant that dominates the striped phase
has the occupation pattern 0000111100000111100001111, which can be
regarded as alternating unit cells of the Pf root configuration 1100
and its particle-hole conjugation 0011. While RH originally interpreted
the striped state as being compressible, in analogy to its higher LL counterparts,
we note their numerical results only indicate broken translation symmetry but
do not provide information on compressibility or charge gap, thus do not contradict
the possibility of an incompressible state here.

In disk geometry, where there is a boundary, the agreement of the
ground state to the Pf or APf is not as good as in a boundaryless
geometry in terms of wave function overlap. However, the present
authors and co-workers showed that a robust Pf state can be identified
by the total angular momentum, edge excitations, and, more
importantly, the possible induction of both Abelian and non-Abelian
quasiholes in a model with realistic confining potential~\cite{wan06}.
When the confining potential becomes much weaker, the ground state can
be identified as the APf state based on the total angular momentum
and the response of the system to the confining potential
tuning~\cite{wan08}. The transition from Pf to APf states driven by background
potential, which breaks particle-hole symmetry, has also been confirmed in the presence of LL
mixing and finite layer thickness~\cite{tylantyler15}.

\begin{figure}
\includegraphics[width=8cm]{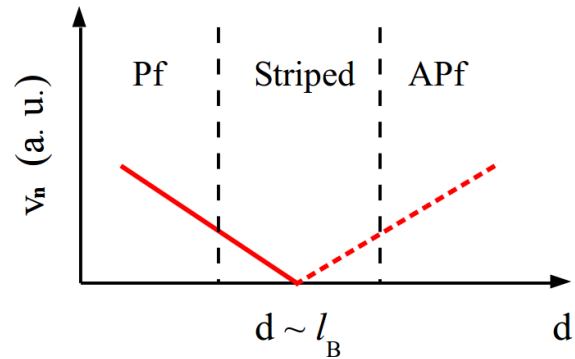}
\caption{\label{fig:phasediagram} (Color online) Phase diagram of the
  Pf, striped, and APf. The striped phase is associated with the smallness
  of the neutral fermionic edge-mode velocity $v_n$, when the energies
  of the Pf and APf states are comparible. The transition can be tuned
  by the confining potential strength, controlled by the setback
  distance $d$ in the numerical model between the two-dimensional
  electron layer and the background charge layer. Note that on the APf
  side we illustrate by a positive $v_n$, but the neutral fermionic
  mode propagates along the opposite direction to its Pf
  counterpart. }
\end{figure}

Interestingly, there is an additional microscopic state {\em
  separating} the Pf and APf phases~\cite{wan08,tylantyler15}, as
illustrated in Fig.~\ref{fig:phasediagram}. This intermediate state was
identified to be a striped phase\cite{wan08}. More recently, Zhang {\it et al.} found that
as the edge confining potential is weakened, the Pf state is destabilized by softening of the (neutral) fermionic edge mode~\cite{zhang14}. This suggests the difference between the resultant striped and Pf states lie in their low-energy {\em neutral} modes. This is clearly consistent with our assessment of the nature of the striped phase. While there is no numerical evidence for it, it is not unreasonable to speculate that the instability of the APf state driven by decreasing $d$ is similarly triggered by softening of neutral mode(s). More detailed discussion on the analysis of disc geometry numerics is presented in the Supplemental Material.

{\it Experimental Consequences.} It has been difficult to understand
that while the theoritical modelings with long-range Coulomb
interaction often show that the Pf or APf state survives in only a
small parameter space\cite{rezayi00}, the FQH effect at $\nu = 5/2$ is observed in
essentially all high mobility samples. With the understanding that the
adjacent striped phase observed numerically is indeed the striped FQH
phase with a charge gap proposed here, the quantized plateau is then
expected to be observable in a much larger parameter space, which is
clearly consistent with experiments.

As discussed earlier, LL mixing induced by Coulomb
interaction breaks the particle-hole symmetry, and can thus drive a
Pf-APf transition~\cite{wang09,bishara09b}, {\em if} they are the only
two competing phases. No such transition, however, has been observed
as the level of LL mixing is varied by changing the electron
density~\cite{nuebler,samkharadze11}, or varying other parameters like
magnetic field~\cite{note}. This can again be easily understood in
terms of the striped quantum Hall phase proposed here, which would
respond to the changing LL mixing and other changes by varying the
relative weight of Pf and APf domains, {\em without} having a phase
transition. The only exception to this is a recent experiment that
claimed to have observed a transition with unusually large Landau
level mixing~\cite{samkharadze13}. In this experiment, a drastic
change of the energy gap dependence on the electron density was
interpreted as the signature of a transition, although there is no
direct method to identify either the Pf state or the APf state. We
believe it is likely that this transition is actually between the
striped state to either the Pf or APf state, with the latter being
stabilized by unusually large LL mixing due to the low
electron density.

More importantly, a distinctive novel property of the striped FQH
phase is the presence of gapless neutral modes (but not the charge
modes) in the bulk, in sharp contrast to all known quantum Hall or
compressible states. A direct experimental observable consequence is
that this electronic system is a bulk thermal conductor and a bulk
charge insulator {\em at the same time}. This contrasts the known
quantum Hall states that transport heat and charge along the edge but
not in the bulk, and compressible states in which the bulk conducts
both charge and heat.  More specifically, in the limit of weak
disorder when we expect the stripes are orderly aligned along a
particular direction, we expect an anisotropic state that conducts
heat along the direction of the stripes but not in the perpendicular
direction, in analogy to the anisotropic conducting state near
half-filled higher
LLs\cite{fogler96,moessner96,lilly99,du99,rhy}. Observation of such
anisotropic heat conduction on a quantum Hall plateau can be
considered definitive evidence of striped quantum Hall state proposed
here.  Stronger disorder may disorder the stripes, rendering the
system an isotropic heat conductor, but still a FQH state.  We note
existence of bulk neutral modes was found in a very recent
experiment~\cite{inoue14}, in which highly sensitive noise measurement
revealed the unexpected heat propagation through incompressible FQH bulk at various filling factors in the lowest
LL. The state proposed in this paper is the first example with such properties, and
it would be interesting to check whether the bulk heat
transport is also present for $\nu = 5/2$.

\begin{figure}
\includegraphics[width=8cm]{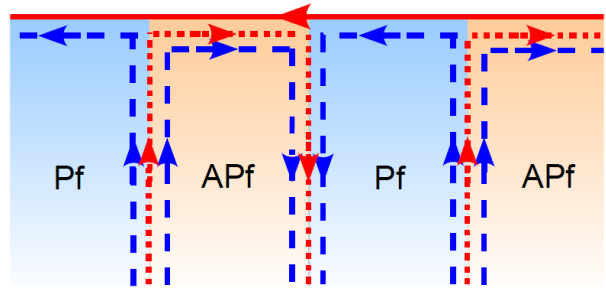}
\caption{\label{fig:edge} (Color online) Illustration of the edge
  structure of the alternating Pf and APf stripes. There is a single
  charged mode (solid red line) that propagates along the edge, while
  the neutral modes (dashed and dotted lines) of the Pf and APf edges
  merge into the bulk neutral modes at their intersections. }
\end{figure}

The striped FQH state has a more complicated edge structure. Depending on how the state terminates, it could either have a Pf edge, APf edge, or more generically, a hybrid between them. The latter would occur when the edge intersects with the domain walls in the bulk, and is illustrated in Fig.~\ref{fig:edge}. In this case there is a {\em single} charged mode that propagates along the edge, while the neutral modes of the Pf and APf edges merge into the bulk neutral modes at their intersections. The rich variety of edge structures may well be responsible for the lack of consistency in the results of existing edge tunneling experiments, and can have profound implications in interferometry experiments. These will be explored in future work.

This work was supported by the 973 Program Project No. 2012CB927404,
NSF-China Grant No. 11174246 (XW), and National Science Foundation Grants
DMR-1442366, DMR-1157490, and the State of Florida (KY).

{\it Note added.} While the paper was being written up we became aware
of a recent preprint~\cite{barkeshli15} on the theory of particle-hole
conjugated composite fermion Fermi liquid states, motivated by a recent experiment
that hinted at a possible particle-hole symmetry breaking near $\nu = 1/2$~\cite{kamburov14}.  In
Ref. \cite{barkeshli15} the {\em possibility} of a phase in which the
charge modes being gapped at an interface between Pf and APf states was
pointed out as a corollary of the authors' detailed analysis of an interface between
particle-hole
conjugated composite fermion Fermi liquid states, although its physical consequences were not discussed.

\newpage

\section{Supplemental Material}

In this section we review the numerical calculation of the FQH state
at $\nu = 5/2$ in the disk geometry with long-range Coulomb
interaction and realistic confining potential. We focus on the
evolution of the ground state as the confining potential becomes
weaker and their compatible theories. Our goal is to show that the
numerical results available is consistent with the existence of a
striped FQH state at $\nu = 5/2$ discussed in the main text.

\subsection{Realistic Models}

\begin{figure}
\includegraphics[width=8cm]{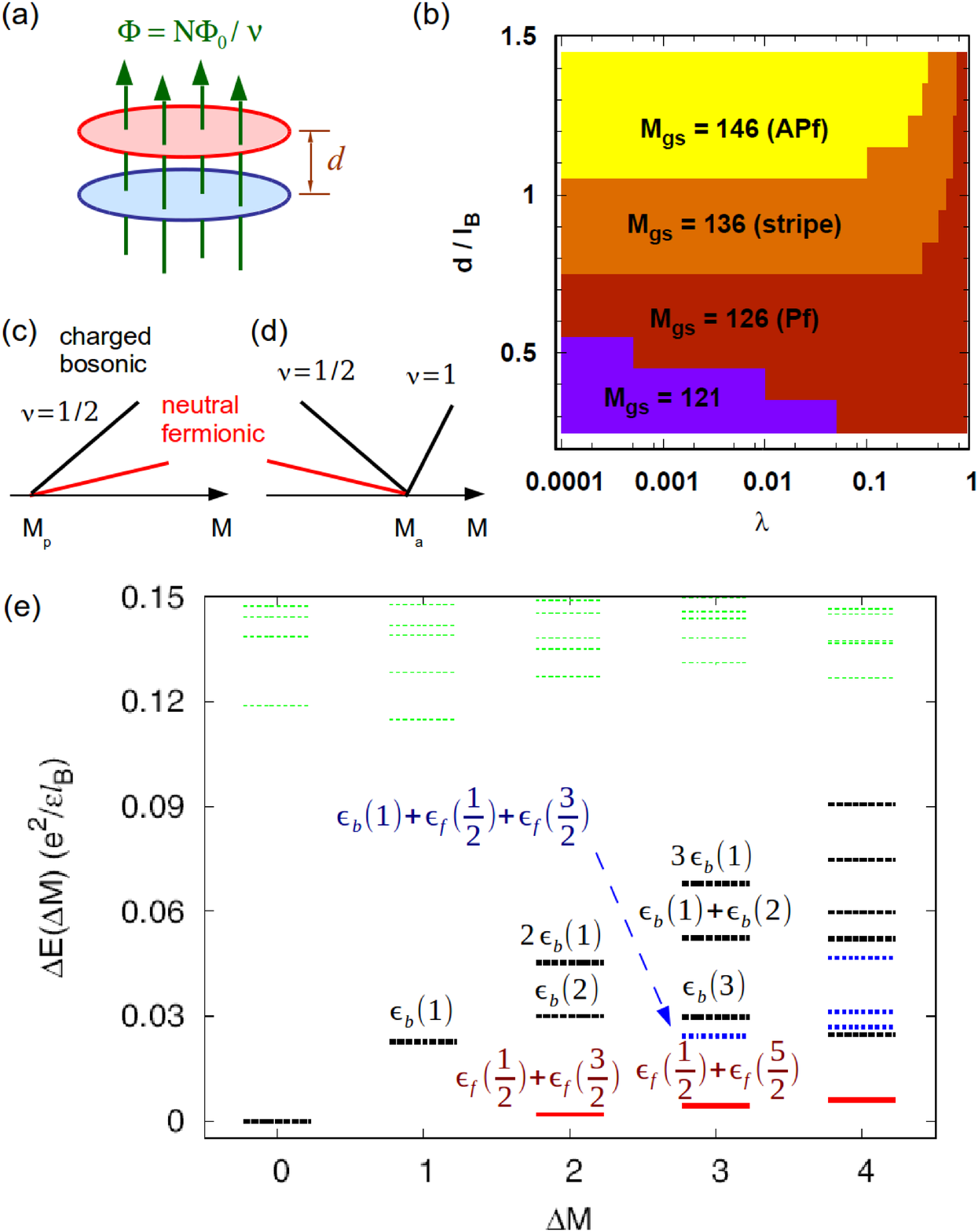}
\caption{\label{fig:sm1} (Color online) (a) Realistic setup of
  two-dimensional electron gas with neutralizing background at a
  setback distance $d$ and uniform magnetic field. (b) Ground state
  angular momentum of 12 valence electrons at $\nu = 5/2$. (c)
  Illustration of the long-wavelength edge modes for the Pf state. (d)
  Illustration of the long-wavelength edge modes for the APf state. (e)
  Identification of the edge states for the 12-electron system at
  $\lambda = 0.5$ three-body potential mixing and $d = 0.6 l_B$. }
\end{figure}

To determine the ground state at $\nu = 5/2$ by numerical calculation
is a formidable task, because one needs to be able to resolve the
subtle energy differences among various competing states, including
the Pf and APf states, in experimentally relevant parameter
space. Ideally, in a realistic model, we may want to consider the
following factors:
\begin{enumerate}
\item Coulomb interaction,
\item neutralizing charge background, which gives rise to the
  confinement of electron near the edge,
\item LL mixing, which induces an effective three-body interaction for
  electrons in the same Landau level,
\item electron layer thickness, which softens the electron-electron
  interaction,
\item electron spin, and
\item disorder.
\end{enumerate}
In the disk geometry, a minimum model that includes factors 1 and 2, is
illustrated in Fig.~\ref{fig:sm1}(a). In the absence of LL mixing, we
only need to consider the valence electrons, or the electrons in the
half-filled first-excited LL (1LL). In this minimum model, the Hamiltonian takes the form
\begin{equation}
\label{eqn:chamiltonian} H_{\rm C} = {1\over 2}\sum_{mnl}V_{mn}^l
c_{m+l}^\dagger c_n^\dagger c_{n+l}c_m +\sum_m U_mc_m^\dagger c_m,
\end{equation}
where $c_m^\dagger$ creates a valence electron with angular momentum
$m$. The Coulomb interaction matrix elements $V_{mn}^l$ in the
symmetric gauge can be expressed conveniently in terms of
pseudopotentials for the 1LL. The confining potential matrix elements
$U_m$ can be calculated by integrating the attraction from the
background charge at a setback distance $d$ away from the electron
layer. In principle, the effect of disorder can be considered by
averaging over the random positional distribution of the ionized
impurities. Practically, one can assume a uniformly distributed
background charge for simplicity. This is the microscopic model used
in Refs.~\cite{wan06sm,wan08sm,zhang14sm}.

High-mobility two-dimensional electron gases are commonly confined in
GaAs quantum wells. The effect of the finite well thickness can be
considered by multiplying to the LL wave functions an additional
factor describing the lowest subband in an infinite rectangular
quantum well. The smearing of the electrons in the quantum well tends
to soften the Coulomb interaction between them. Because the Moore-Read
Pf state is the exact ground state of a three-body interaction, LL
mixing, which generates high-order corrections (including three-body
terms) to the Coulomb interaction due to finite LL splitting, is
important. For numerical work, one can include the corrections in the
form of two-body and three-body
pseudopotentials~\cite{peterson13sm}. The minimal model, now with LL
mixing and finite well thickness, is considered in
Ref.~\cite{tylantyler15sm}. Even before the perturbative LL mixing
corrections were worked out, numerical models that include the
Coulomb interaction with three-body interaction on phenomenological
ground had been used~\cite{wan06sm}.

In all of these numerical works, electrons are considered to be spin
polarized. This assumption is supported by the density matrix
renormalization group calculation in the spherical
geometry~\cite{feiguin09sm} in the absence of LL mixing, and the
truncated Hilbert space calculation in the torus geometry in the
presence of LL mixing and finite layer thickness~\cite{rezayi11sm}.

\subsection{Ground States and Edge Modes}

The total angular momentum $M$ of the 12-electron ground state of the
minimum model with an extra phenomenological three-body interaction
\begin{equation}
\label{eqn:mixedhamiltonian} H = (1 - \lambda)
H_{\rm C} + \lambda H_{3B}
\end{equation}
is mapped in Fig.~\ref{fig:sm1}(b) as a function of both the setback
distance $d$, in units of magnetic length $l_B$, and the three-body
content $\lambda$, as was previously reported in Ref.~\cite{wan08sm}.
The ground state with $M = 126$ has been unambiguously identified as
the Pf phase for the following reasons.
\begin{enumerate}
\item Its momentum $M$ matches that of the 12-electron Moore-Read Pf
  state $M_p$~\cite{wan06sm}.
\item Increasing three-body interaction strength enhances the
  state~\cite{wan08sm}.
\item Its edge spectrum matches that of the Pf state, with one branch
  of bosonic mode and one branch of fermionic mode~\cite{wan06sm,wan08sm}.
\item A local potential trap with different strength can stabilize
  charge $e/2$ quasihole and charge $e/4$ quasihole. In the latter
  case, the change of the edge-mode counting confirms that the charge
  $e/4$ quasihole is a non-Abelian anyon~\cite{wan06sm}.
\end{enumerate}
Note that we intentionally downplay the role of the wave function
overlap that is often used in a closed geometry as an indicator; the
existence of the edge significantly lowers the overlap, but it
provides extra information in return through the bulk-edge
correspondence as we have already seen. In particular, the energetics
of the edge spectrum can be used to extract information on the
dispersion relation of the charged bosonic mode and the neutral
fermionic mode, as illustrated in Fig.~\ref{fig:sm1}(e) for $d = 0.6
l_B$ and $\lambda = 0.5$ (see Refs.~\cite{wan08sm,zhang14sm} for more
detailed illustrations). As a result, the fermionic edge mode is found
to have significantly smaller velocity than that of the bosonic mode
~\cite{wan08sm}. Both are propagating along the same direction, so we
can represent the edge spectrum by two straight lines with different
slopes in the cartoon plot Fig.~\ref{fig:sm1}(c), which is meant to
illustrate how energies of the bosonic or fermionic edge excitations
grows with momentum around the value $M_p$ for the Pf state in the
long-wavelength limit.

For comparison, the ground state with $M = 146$ is identified to be the
APf state for the following reasons.
\begin{enumerate}
\item Its momentum $M$ matches that of the 12-electron APf
  state $M_a$~\cite{wan06sm}.
\item Increasing three-body interaction strength suppresses the
  state~\cite{wan08sm}.
\item The state is robust with the inclusion of two extra
  orbitals~\cite{wan08sm}.
\item Smoother confining potential can induce a state that is
  consistent with the APf state with a charge $e/4$
  quasihole~\cite{wan08sm}.
\item There exists a low-energy excitation at $\Delta M = -2$,
  corresponding to the smallest-momentum fermionic edge mode. As $d$
  decreases, its energy decreases toward zero; for comparison, as $d$
  increases in the Pf phase, the energy of the lowest $\Delta M = 2$
  state also decreases toward zero.
\end{enumerate}
No further exploration on the APf state was carried out in Ref. \cite{wan08sm} due to
the complicated edge structure, as illustrated by the cartoon in
Fig.~\ref{fig:sm1}(d).

Reason 5 in the APf case hints a transition from the Pf state to the
APf state around $d = l_B$ and an approximate mirror symmetry between
the two phases around the transition. However, numerical calculation
found that there is an intermediate state ($M = 136$), which appears
to be of stripe nature~\cite{wan08sm}. The presence of the
intermediate state is robust even when LL mixing and finite layer
thickness are considered~\cite{tylantyler15sm}.

\subsection{The Transition from Pf to APf}

It is interesting to ask what emerges to be the ground state when one
tunes the confining potential of the Pf state weaker and weaker. We
already pointed out in the previous subsection that the APf state is
not emerging directly. Two simplest possibilities, then, are as
follows.
\begin{enumerate}
\item The Pf state with a quasihole in the bulk. Due to the rotational
  symmetry, such a quasihole can only appear at the center. However,
  the total angular momentum of the observed state ($M = 136$) matches
  neither the case of an Abelian $e/2$ quasihole ($M = 138$) nor the
  case of a non-Abelian $e/4$ quasihole ($M = 132$)~\cite{wan06sm}.
\item Bosonic edge reconstructed Pf state. Bosonic edge mode at small
  $\Delta M$ shows signature of bending down, as for the Laughlin
  state~\cite{wan02sm,wan03sm}. If this persists to larger $\Delta M$,
  the chiral boson edge mode becomes soft and the bosons
  condense. This is the microscopic interpretation of the edge
  reconstruction, which means that we expect to
  see charge density modulation around the edge. The fermionic mode is
  not necessary to be involved.
\end{enumerate}

\begin{figure}
\includegraphics[width=7cm]{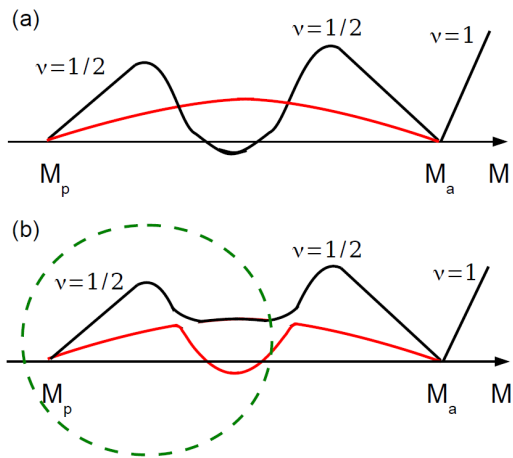}
\caption{\label{fig:sm2} (Color online) Possible profile of the
  low-energy states at $d \sim l_B$. The ground state with the angular
  momentum in between the Pf and APf values has a lower energy than
  either the Pf or APf state and is possibly of either (a) bosonic
  nature or (b) fermionic nature. The red lines schematically
  represent the lowest fermionic excitations, while the black ones the
  lowest bosonic ones. In (a) the bosonic mode reconstructs. The
  energy dispersion curves of the edge modes obtained by Zhang {\it et
    al.}~\cite{zhang14sm}, on the other hand, can reproduce the
  profile in the circled area in (b).}
\end{figure}

If the bosonic edge reconstruction were the origin of the striped
phase, we would expect to observe a crossing of the bosonic and
fermionic modes at a sufficiently large $\Delta M$, because at small
$\Delta M$ the bosonic mode has a larger velocity. Note that the new
ground state has an angular momentum in between those of the Pf and
the APf states. We can schematically illustrate the profile of the
low-energy states, including the edge excitations of the Pf and APf
states, as in Fig.~\ref{fig:sm2}(a).

Very recently, Zhang {\it et al.}  studied the possibility of edge
reconstruction at $\nu = 5/2$~\cite{zhang14sm}. In the minimum model
they obtained the edge states of the Pf state by diagonalization in
the restricted Pf basis, and resolved them by applying a density
operator for bosonic edge excitations and trial wave functions for
fermionic edge excitations, in addition to the self-consistency
conditions of the edge-mode dispersions. They found that the bosonic
edge mode persists to have a higher energy scale than the fermionic
mode, up to the capability of their resolution~\cite{zhang14sm}. Their
findings indicate that the usual bosonic edge reconstruction is less
likely to be responsible for the edge instability of the Pf
state. Instead, they proposed a scenario of Majorana-driven edge
reconstruction~\cite{zhang14sm}; this requires that the dispersion of
the fermionic mode develop a roton-like structure, which is pushed down,
presumably, by the bending down bosonic mode. Therefore, the profile
of the low-energy excitations resembles the sketch in
Fig.~\ref{fig:sm2}(b), in the green circle of which the discovery of
Zhang {\it et al.}~\cite{zhang14sm} is illustrated. We note the restricted basis used
there limits the possible types of instabilities the Pf state (as we discuss below); but 
this work suggests that the instabilities are likely triggered by softening of
{\em neutral} degree of freedom.

\begin{figure}
\includegraphics[width=8cm]{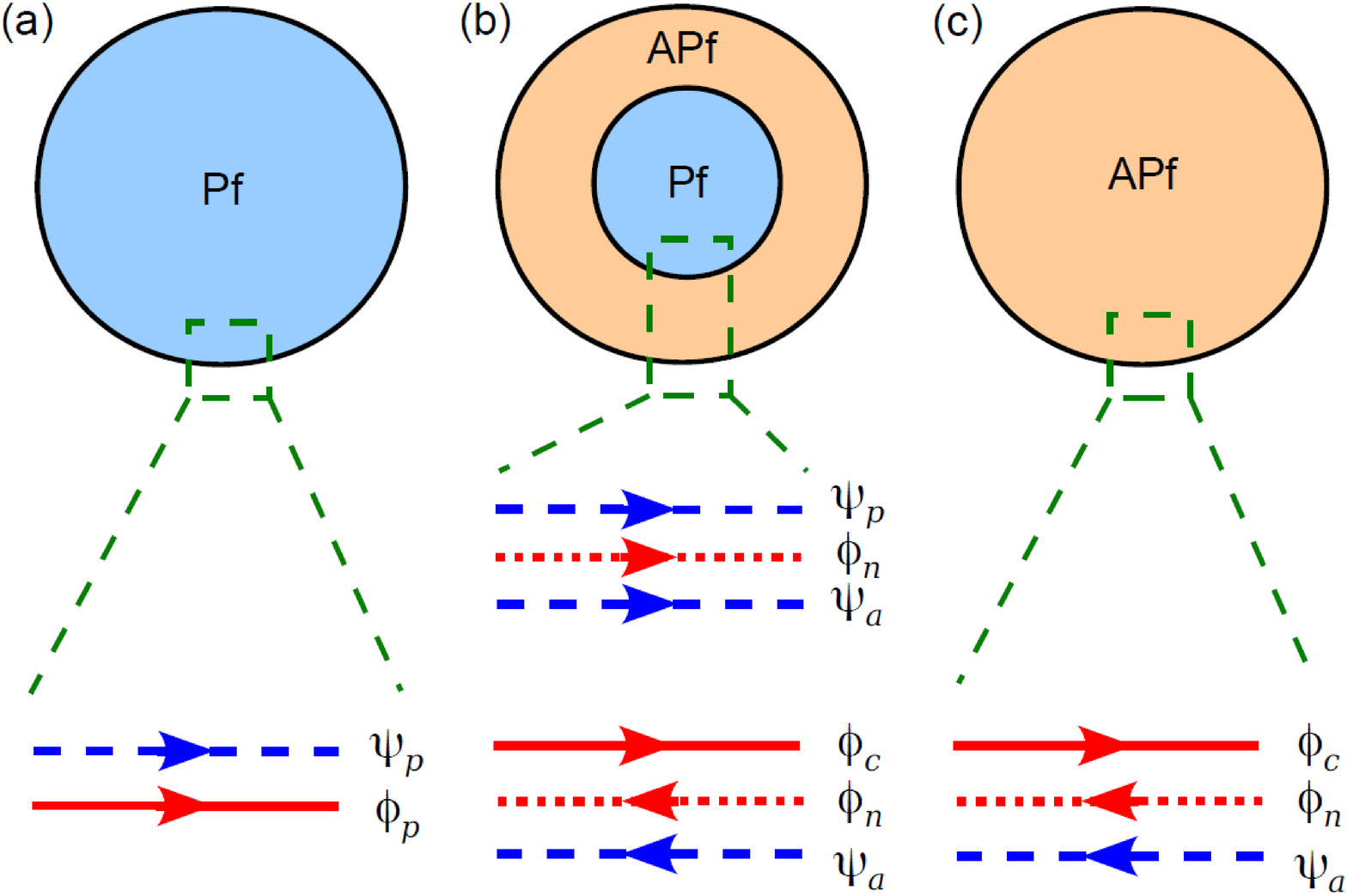}
\caption{\label{fig:sm3} (Color online) Evolution of a FQH droplet at
  $\nu = 5/2$. (a) Homogeneous Pf state is stable at a sharp confining
  potential. There are one charged bosonic mode and one neutral
  fermionic mode at the edge propogating along the same direction. (b)
  Pf-APf striped phase, driven by edge reconstruction, emerges at
  smoother confining potential. At the Pf-APf interface the bosonic
  modes gapped out and there are only neutral modes propagating along
  the same direction. (c) Homogeneous APf state is stable at a smooth
  confining potential. In (b) and (c) we assume the APf edge modes are
  strongly coupled such that there are only one charge mode, one
  counterpropogating neutral bosonic mode, and one counterpropogating
  neutral fermionic mode.}
\end{figure}

Proposed in the present work is an instability appearing at the edge, 
but the edge piece is an APf stripe. Naturally, higher
electron density at the APf edge, as oppose to the Pf edge, is
energetically favorable where the confining potential is smooth. The
scenario is illustrated in Fig.~\ref{fig:sm3} in disk geometry. When
the confining potential is sharp, the Pf state is favored and there
are one charged bosonic mode and one neutral fermionic mode at the
edge [Fig.~\ref{fig:sm3}(a)]. Smoother confining potential drives edge
reconstruction and induces an APf ring around the Pf bulk
[Fig.~\ref{fig:sm3}(b)]. At the Pf-APf interface the bosonic modes are
gapped out and there are neutral modes only as discussed in the main
text; this is also consistent with the numerical results in Zhang {\it
  et al.}~\cite{zhang14sm}. The outer edge is an APf edge; here we
illustrate with the case in which the two counter-propagating charge
modes are coupled, leading to one charge mode and one neutral mode.
To see that this transition is due to edge reconstruction, we can
compare the low-energy modes in terms of chiral central charges for
left- and right-going modes. The chiral central charges are $(c_+,
c_-) = (1.5, 0)$ for the Pf state, and $(3, 1.5)$ for the striped
state. Note the difference $c_+ - c_- = 1.5$ is the same for them,
allowing an edge instability to connect them.  As the confining
potential smooths further, the Pf core shrinks to the center and
eventually an homogeneous APf state is left
[Fig.~\ref{fig:sm3}(c)]. Therefore, the existence and evolution
between the three phases are consistent with all available numerical
results in disk geometry. We note in such a small system sizes there 
is only one interface between Pf and APf state in the striped phase.
For large systems we expect many interfaces that are separated by distance(s)
of order several magnetic length, thus forming a true bulk phase as discussed
in the main text.


\begin{thebibliography}{99}
\bibitem{note} The Pf and APf states are expected to have different bulk excitation spectra in the absence of particle-hole symmetry, thus a transition between them can occur when quasiparticle or quasihole densities vary due to change of magnetic field with fixed electron density. No such transition has been observed.
\end{thebibliography}
\end{document}